\documentclass[fleqn,twoside]{article}
\usepackage[headings]{espcrc2}

\readRCS
$Id: espcrc2.tex,v 1.2 2004/02/24 11:22:11 spepping Exp $
\usepackage{graphicx}
\usepackage{graphics,epsfig}
\usepackage{cite} 

\newcommand{\mathbold}[1]{\mbox{\boldmath $\bf #1$}}

\newcommand{\sw}{s_W}
\newcommand{\MZ}{{M_Z}}
\newcommand{\CZb}{\cos 2\beta\hspace{1mm}}
\newcommand{\mq}{m_{q}}
\newcommand{\tb}[1][]{\ensuremath{\tan^{#1}\!\beta}}
\newcommand{\nn}{\nonumber}

\newcommand{\be}{\begin{equation}}
\newcommand{\ee}{\end{equation}}

\def\lsim{\mathrel{\raise.3ex\hbox{$<$\kern-.75em\lower1ex\hbox{$\sim$}}}}
\def\gsim{\mathrel{\raise.3ex\hbox{$>$\kern-.75em\lower1ex\hbox{$\sim$}}}}
\def\ifmath#1{\relax\ifmmode #1\else $#1$\fi}
\def\ls#1{\ifmath{_{\lower1.5pt\hbox{$\scriptstyle #1$}}}}

\newenvironment{Eqnarray}%
     {\arraycolsep 0.14em\begin{eqnarray}}{\end{eqnarray}}

\title{Quantum corrections to the MSSM $h^0 b \bar b$ vertex: Decoupling limit}

\author{Howard E. Haber\address{Santa Cruz Institute for Particle
   Physics, University of California, Santa Cruz, CA 95064, USA},
        Heather E. Logan\address{Ottawa-Carleton Institute for Physics, 
Dept. of Physics, Carleton University,
1125 Colonel By Drive, Ottawa, ON K1S 5B6, Canada},
        \underline{Siannah Pe\~naranda}\address{CERN-TH Division, 
Department of Physics, CH-1211 Geneva 23, Switzerland}%
        \thanks{The work of S.P. has been supported by the 
European Union under contract No.~MEIF-CT-2003-500030},
        and
        David Temes\address{INFN, Laboratori Nazionali di Frascati, 
I-00044 Frascati, Italy}}

\runtitle{Quantum corrections to the MSSM $h^0 b \bar b$ vertex: 
Decoupling limit}
\runauthor{H.E. Haber, H.E. Logan, S. Pe\~naranda, D. Temes}

\begin{document}

\begin{abstract}
We consider the leading one-loop Yukawa-coupling 
corrections to the $h^0 b\bar{b}$ coupling at ${\cal O} (m_t^2)$ 
in the MSSM in the decoupling limit. 
The decoupling behavior of the corrections from the various MSSM sectors
is analyzed in the case of having some or all of the supersymmetric mass
parameters and/or  the CP-odd Higgs mass large
as compared to the electroweak scale. 
\end{abstract}

\maketitle

\section{Introduction}
It is well known that the tree-level couplings of the lightest
Minimal Supersymmetric Standard Model (MSSM)
 Higgs boson ($h^0$) to fermion pairs and gauge bosons
tend to their Standard Model (SM) values
in the decoupling limit, $M_A \gg M_Z$ \cite{decoupling}.
As a consequence of this decoupling, distinguishing
the lightest MSSM Higgs boson in the large $M_A$
limit from the Higgs boson of the SM will be very
difficult. Our aim is to determine the nature of the decoupling limit
at one-loop for the couplings of $h^0$ to SM particles.
If some non-decoupling behavior of supersymmetric (SUSY) 
particles is found, it will provide a
clear signal for some low energy observables, even if $M_{SUSY} \sim {\cal{O}}
({\rm TeV})$. 

In this paper, we focus on the
$h^0$ coupling to $b\bar b$.  This coupling determines the partial width
of $h^0 \to b\bar b$, which is by far the dominant decay mode of $h^0$ in most
of the MSSM parameter space.  Therefore, 
accurate knowledge of the $h^0 b \bar b$
coupling is very important for Higgs boson searches. In particular, we
study the ${\cal O} (m_t^2)$ Yukawa coupling MSSM radiative corrections
to the $h^0 b \bar b$ vertex at one loop level, and we explore 
their behavior in the decoupling limit.  
A detailed discussion of the
SUSY-QCD corrections that arise from gluino and bottom-squark
(sbottom) exchange have been previously given in~\cite{nos}. 
It has been shown that in the decoupling limit of both 
large SUSY mass parameters and large CP-odd Higgs mass,
the $h^0\rightarrow b \bar b$ decay width approaches its SM 
value at one loop, with the onset of decoupling delayed for large
$\tan\beta$ values. However, this decoupling does not occur if just the  
SUSY mass parameters are taken large.

The full diagrammatic formula for the
on-shell EW (electroweak)-Yukawa corrections 
to the $h^0 b \bar b$ coupling will be presented in~\cite{newhbb}.  
Here we summarized the results obtained in this paper.
In Section~\ref{sec:masses} we briefly review the decoupling limit in 
the Higgs sector and the SUSY sector of the MSSM.
The ${\cal O} (m_t^2)$ Yukawa corrections to the $h^0 b \bar b$
coupling are presented in Section~\ref{sec:renormalization}.
Some details of the renormalization procedure and a discussion of the
decoupling properties of these corrections are included in this
section. Analytical and numerical results are collected in 
section~\ref{sec:analytic}. We conclude in
Section~\ref{sec:conclusions}.

\section{Decoupling limit in the MSSM}
\label{sec:masses}

The properties of the MSSM Higgs sector at the
tree-level are determined by just two free parameters,
conventionally chosen as the mass of
the CP-odd neutral Higgs boson ($A^0$), $M_A$, and the ratio of the
vacuum expectation values (\textit{vev}s) of each doublet,
$\tb=v_2/v_1$~\cite{HHG}.

The decoupling limit in the MSSM is defined by considering the 
parameter regime where $M_A\gg M_Z$. 
In this limit, the expressions for the
Higgs masses and mixing angle simplify~\cite{decoupling} and 
two consequences are immediately apparent.
First, $M_A\simeq M_{H^0}\simeq M_{H^\pm}$,
up to corrections of ${\mathcal O}(M_Z^2/M_A)$, and
$M_{h^0} \simeq M_Z |\cos 2\beta|$.  Second,
$\cos(\beta-\alpha)=0$ up to corrections of ${\cal O}(M_Z^2/M_A^2)$.
Consequently, the effective Higgs sector consists only of one light
CP-even Higgs boson, $h^0$, whose couplings to SM particles
are indistinguishable from those of the SM Higgs boson.
When radiative corrections to the CP-even Higgs mass-squared matrix
are taken into account, the upper bound on $M_{h^0}$
increases substantially to $M_{h^0} \lsim 135$ GeV~\cite{hradcorr}.

Summarizing the parameters of the squark sector, 
the tree-level squark squared-mass matrix is:
\begin{equation}
    {\hat M}_{\tilde q}^2 \equiv\left(\begin{array}{cc}  
 M_{\tilde L_q}^2 & m_q X_q  \\ m_q X_q & M_{\tilde R_q}^2   
\end{array} \right) \,\,,\,\,\,\,\,\,{\footnotesize{q\equiv t,b}}
\label{eq:sbottommatrix}
\end{equation}
with,
\begin{Eqnarray}
  M_{\tilde L_q}^2 &=& M_{\tilde Q}^2 + \mq^2 + 
                     \CZb \MZ^2 (T_3^q - Q_q \sw^2) \nn \\
  M_{\tilde R_q}^2 &=& M_{\tilde U\,, \tilde D}^2 + \mq^2 + 
                     \CZb \MZ^2 Q_q \sw^2 \nn \\
 X_t &=& A_t- \mu \cot{\beta}\,,\,\,\,\,\,\,
 X_b = A_b- \mu \tan{\beta}\,,
    \label{eq:sbottomparams}
\end{Eqnarray}%
and $s_W\equiv \sin\theta_W$.  The parameters
$M_{\tilde Q}$ and $M_{\tilde U, \tilde D}$ are the soft-SUSY-breaking masses, 
$A_t$ is a soft-SUSY-breaking trilinear
coupling and $\mu$ is the bilinear coupling of the two Higgs doublet
superfields. 

In order to get heavy squarks, we need to choose 
large values for the appropriate
soft SUSY breaking parameters and the $\mu$-parameter. Since 
we are interested here in the limiting situation where the whole 
SUSY spectrum is heavier than the electroweak scale, we have made
the following assumptions (see ref.~\cite{newhbb} for more details),
\begin{Eqnarray}
M_{{\scriptscriptstyle {\tilde Q},{\tilde U},{\tilde D}}} 
\sim M_{\tilde g} 
\sim \mu  \sim A_{t,b}  \sim M_{SUSY}\gg M_Z,
\label{eq:limitq}
\end{Eqnarray}%
where $M_{SUSY}$ represents generically a common large SUSY mass scale.
In addition, 
we have considered two extreme cases, maximal and minimal mixing,  
 which imply certain constraints on the squark mass differences:  
A. Close to maximal mixing ($\theta_{\tilde q} \sim \pm 45^{\circ}$):
$|M_L^2-M_R^2|\ll m_q X_q \Rightarrow 
|M_{\tilde q_1}^2-M_{\tilde q_2}^2|\ll 
|M_{\tilde q_1}^2+M_{\tilde q_2}^2|$, and 
B. Close to minimal mixing ($\theta_{\tilde q} \sim 0^{\circ}$):
$|M_L^2-M_R^2|\gg m_q X_q \Rightarrow 
|M_{\tilde q_1}^2-M_{\tilde q_2}^2|\sim
{\cal O}|M_{\tilde q_1}^2+M_{\tilde q_2}^2|$.
Eq.~(\ref{eq:limitq}) also implies that the gluino is heavy.

Finally, the chargino mass matrix is
given by,
\begin{Eqnarray}
{\hat M}_{{\tilde \chi}^{\pm}}=\left(\begin{array}{cc}
M_{2} & \sqrt{2} m_{\scriptscriptstyle W} \sin{\beta} 
\\ \sqrt{2} m_{\scriptscriptstyle W} \cos{\beta} & \mu \,
\end{array} \right)\,\,\,,
\end{Eqnarray}%
and in order to get heavy charginos we consider the limit:
$\,M_{SUSY} \sim \mu  \sim M_{2} \gg M_Z $. 


\section{$\mathbold{\mathcal{O} ({\it m}_{\it t}^2)}$ Yukawa corrections 
to $\mathbold{{\it h}^0\, \rightarrow  {\it b} \,\bar {\it b}}$}
\label{sec:renormalization}

Here we present the one-loop corrections to the partial
decay width $\Gamma ( h^0 \to b \bar b)$.  We will then explore
the decoupling behavior of these corrections for large SUSY masses, 
$M_{SUSY}$, and/or large $M_A$. Both numerical and analytical results will be 
presented elsewhere~\cite{newhbb}.

The tree-level $h^0 b \bar b$ coupling is given by
\begin{equation} \label{eq.hbbtree}
    g_{hbb} = \frac{gm_b \sin\alpha}{2M_W \cos\beta}\,.
\end{equation}
In lowest order this Higgs-fermion vertex represents the Yukawa coupling
proportional to the fermion mass $m_f=m_b$. 
Note that in the limit of large $M_A$, $\sin\alpha \to - \cos\beta$
and $g_{hbb}$ tends to the SM coupling, $g_{hbb}^{SM}=-{gm_b}/(2M_W)$.

The vertex functions obtained from the set of one-loop diagrams
are in general UV-divergent.
For finite one-particle irreducible (1PI) Green functions and 
physical observables,
renormalization has to be performed by adding appropriate counterterms.
We follow the conventions given in~\cite{dabelsteinrenorm} 
for the renormalization procedure. 

\subsection{The ${\mathbold{\alpha}}_{{\mbox{eff}}}$-approximation}
\label{sec:renorcond}

The radiatively-corrected $h^0 b \bar b$ coupling depends on the
mixing angle $\alpha$.  At tree-level, $\alpha$ is determined
by fixing $\tan\beta$ and $M_A$.  At one-loop order, there are no
$\mathcal{O}(\alpha_s)$ corrections to this mixing angle~\cite{nos}. However, 
once one-loop SUSY-EW effects are included, 
the one-loop radiative corrections to $\alpha$
must be taken into account~\cite{cmwpheno}.
It turns out that the dominant contributions 
to the Higgs boson self-energies can be obtained by setting $q^2=0$. 
Consequently, these corrections can be absorbed into a redefinition of the
CP-even neutral Higgs mixing angle $\alpha$.  This is the so-called
${\alpha}_{{\mbox{eff}}}$-approximation.  

More explicitly,
we approximate the renormalized Higgs boson self-energies by 
$\hat{\Sigma}(q^2)\simeq\hat{\Sigma}(0) \equiv \hat{\Sigma}$.
Consequently, the correction to the mixing
angle, $\Delta \alpha$, is related to the
renormalized self-energies and masses.
Neglecting terms beyond one-loop level, one obtains
\begin{equation}
\tan \,\Delta \alpha = \frac{\hat \Sigma_{h^0 H^0}}{M_{h_o}^2 - M_{H_o}^2}\,,
\label{tandeltaalpha}
\end{equation}
where $\hat \Sigma_{h^0 H^0}$ is the renormalized $h^0$--$H^0$ mixing 
propagator given in~\cite{dabelsteinrenorm}.
We can then absorb the contributions to the $h^0$--$H^0$
mixing, due to the squarks/chargino loops, in the redefinition of the 
effective mixing angle $\alpha$. 
In this approximation one deduces that 
$\mathcal{Z}_R^{h^0} \approx 1$. 

\subsection{Analytic and numerical results}
\label{sec:analytic}

\begin{figure}
\begin{center}
\epsfig{figure=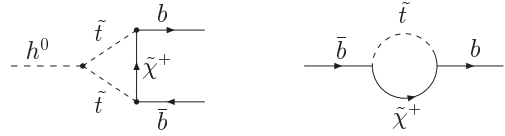,width=7.4cm}\vspace*{-1cm}
\caption{Feynman diagrams for the ${\mathcal{O}}(m_t^2)$
corrections to the $h^0 b \bar b$ coupling.}
\label{fig:fd}
\end{center}\vspace*{-1cm}
\end{figure}

In this section we present explicit results for the
${\mathcal{O}}(m_t^2)$ EW-Yukawa correction to the $h^0 b \bar b$ vertex. 
The leading Yukawa contribution to the $h^0 b \bar{b}$ coupling
arises from diagrams involving the
exchange of virtual top-squarks, as shown in Fig.~\ref{fig:fd}.

At the one-loop level and ${\mathcal{O}}(m_t^2)$, 
the $h^0 b \bar{b}$ coupling can be written as,
\begin{equation}
    \bar{\Gamma}(h^0 \to b \bar b) = \Gamma(h^0 \to b \bar b)
    (1 + 2 \Delta_{{\rm SUSYEW}})\,,
\end{equation}
where $\bar{\Gamma}$ is the one-loop partial width and $\Gamma$
is the tree-level partial width as in~(\ref{eq.hbbtree}).
$\Delta_{{\rm SUSYEW}}$ denotes the ${\mathcal{O}}(m_t^2)$ radiative 
corrections to this vertex as given in Fig.\ \ref{fig:fd},
\begin{equation}
\Delta_{{\rm SUSYEW}}=\Delta_{{\rm SUSYEW}}^{\rm loops}+
\Delta_{{\rm SUSYEW}}^{\rm CT}\,.
\end{equation}
The triangle diagram, with the exchange of stops and charginos, 
contributes to $\Delta_{{\rm SUSYEW}}^{\rm loops}$, whereas the bottom 
self-energy diagram contributes to the contribution of the counterterms,  
$\Delta_{{\rm SUSYEW}}^{\rm CT}$.
We have checked that other triangle contributions 
(with one top-squark and two charginos) and neutralinos contributions 
are subleading diagrams; hence, these diagrams are not included here. 
Note that our results are in agreement with the corresponding results 
of~\cite{Dabelstein}. 

Expansions in inverse powers of SUSY masses are performed in order to
examine the decoupling behavior when the SUSY masses 
are large compared to $M_Z$. Concretely, 
we perform expansions of the loop integrals and mixing angles 
for $M_{SUSY}\gg m_{EW}$ and isolate terms of
${\cal O}\left({m_{EW}^2}/{M_{SUSY}^2})^n\right)$. Thus, by defining 
\begin{equation}
\tilde M_S^2 \equiv 
\frac{1}{2}(M_{\tilde t_1}^2 + M_{\tilde t_2}^2)\,,\,\,\quad
 R \equiv M_{\tilde \chi^{+}}/\tilde M_S\,,
\end{equation}
and including the leading $\mathcal{O}(1)$ 
terms (\textit{i.e.}, $n=0$)
in the expansion,, we obtain 
the following result
for the maximal mixing case, $\theta_{\tilde t}\sim \pm 45 ^{\circ}$:
\begin{Eqnarray}%
        && \hspace*{-0.1cm}{\Delta_{{\rm SUSYEW}}}=
\frac{g^2}{64\pi^2 m_W^2} \frac{1}{\sin^2 \beta}\,m_t^2\,\times \nn\\
        && \hspace*{-0.1cm}\left\{\frac{-\mu A_t}{\tilde M_S^2}
        \left(\tan\beta + \cot\alpha \right) f_1(R)  + 
        {\mathcal O}\left(\frac{m_{EW}^2}{M_{SUSY}^2}\right)\right\},\nn\\
\label{eq:amalitical}
\end{Eqnarray}%
 where the functions $f_i(R)$ are defined in ref.~\cite{nos} and 
 have been normalized as $f_i(1)=1$.

Notice that the first term in~(\ref{eq:amalitical}) is the dominant one in the
 limit of large $M_{SUSY}$ [eq.~(\ref{eq:limitq})] and does not vanish 
 in the asymptotic limit of infinitely large 
 ${\tilde M_S}$, $\mu$ and $A_t$. Therefore this term gives
  a non-decoupling SUSY contribution to the $\Gamma(h^o \to b \bar b)$ partial
 width which can be of phenomenological interest. Moreover, since this term 
 is enhanced at large $\tan \beta$ it can provide important corrections to the
 $h^o \to b \bar b$ total width, even for a very heavy SUSY
 spectrum. The sign of these corrections 
 are fixed by the sign of $\mu \,A_t$. We find similar results for the 
minimal mixing case, $\theta_{\tilde b}\sim \pm 0 ^{\circ}$~\cite{newhbb}. 

From this result, we conclude that there is no decoupling of stops 
and charginos in the limit of large SUSY mass parameters for fixed
$M_A$. Similar results have been found for the SUSY-QCD corrections~\cite{nos}.
In contrast, most numerical studies done so far on this subject 
indicate decoupling of heavy SUSY particles from SM physics. 
  How do we then recover
 decoupling of the heavy MSSM spectra from the SM low energy physics? The
 answer to this question relies in the fact that in order to converge to SM
 predictions we need to consider not just a heavy SUSY spectra but 
also a heavy Higgs sector. That is, besides large $M_{SUSY}$, 
the condition of large $M_A$ 
 is also needed. Thus, if $M_A \gg M_Z$, one can easily derive:
\begin{equation}
    \cot\alpha = -\tan\beta - \frac{2M_Z^2}{M_A^2} \tan\beta \cos 2\beta
    + \mathcal{O}\left(\frac{M_Z^4}{M_A^4}\right).
\label{eq:cotalphaexpansion}
\end{equation}
By substituting this into~(\ref{eq:amalitical}) we see that the non-decoupling
 terms cancel out and we end up with
\begin{Eqnarray}
        && \hspace*{-0.1cm}\Delta_{{\rm SUSYEW}}
        = \frac{g^2}{32\pi^2 m_W^2} \frac{1}{\sin^2
        \beta}\,\,m_t^2\,\times \nn\\
        && \hspace*{-0.2cm}\left\{ \frac{-{\mu A_t}}{\tilde M_S^2}
        f_1(R)\, \tan\beta \cos 2\beta \,\frac{m_Z^2}{m_A^2}  
        + {\mathcal O} 
        \left(\frac{m_{EW}^2}{M_{SUSY}^2}\right) \right\}\nn\\
\end{Eqnarray}%
 which clearly vanishes in the asymptotic limit of 
 $M_{\scriptscriptstyle {SUSY}}$ 
 and  $M_{\scriptscriptstyle {A}} \rightarrow \infty$. 
Therefore, we get decoupling if and only if both 
$M_{\scriptscriptstyle {SUSY}}$ and $M_{\scriptscriptstyle {A}}$ are large. 

\begin{figure}[t]
\vspace*{0.2cm}
\epsfig{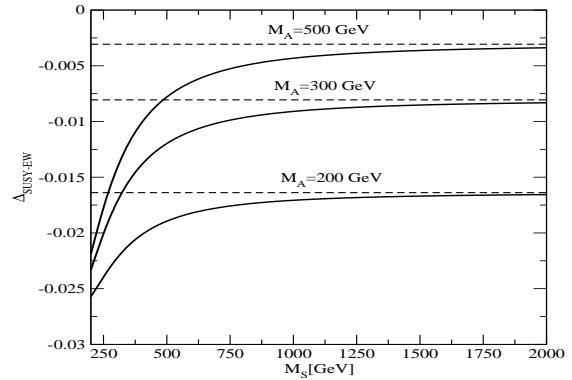}\vspace*{-0.8cm}
\caption{$\Delta_{{\rm{SUSYEW}}}$ as a function of $M_{SUSY}$
for $M_A = 200$, 300, and 500 GeV and $\tb=30$.}\vspace*{-0.4cm}
\label{fig:MSMA}
\end{figure}

\begin{figure}[thb!]
\vspace*{0.1cm}
\epsfig{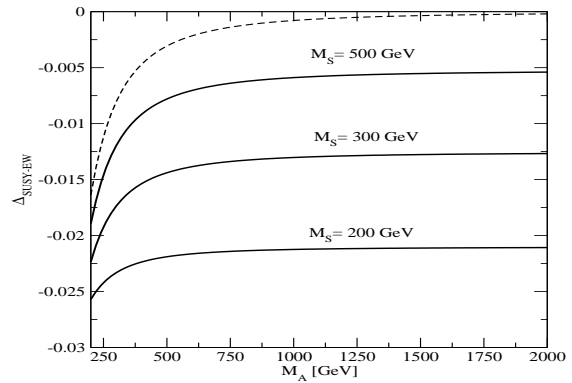}\vspace*{-0.8cm}
\caption{$\Delta_{{\rm{SUSYEW}}}$ as a function of $M_A$
for $M_{SUSY} = 200$, 300, and 500 GeV and $\tb=30$.}\vspace*{-0.6cm}
\label{fig:MAMS}
\end{figure}

The above non-decoupling behavior is shown numerically in 
Figs.\ \ref{fig:MSMA} and\ \ref{fig:MAMS} for $\tb=30$ and SUSY 
parameters as defined in~(\ref{eq:limitq}). We show the dependence
of $\Delta_{{\rm SUSYEW}}$ on $M_{SUSY}$ (Fig.\ \ref{fig:MSMA}) and $M_A$ 
(Fig.\ \ref{fig:MAMS}).  
Clearly, in the limit of large $M_{SUSY}$, $\Delta_{{\rm SUSYEW}}$
tends to a non-vanishing constant, and this constant tends to zero
in the large $M_A$ limit. 
Similarly, in the limit of large $M_A$, $\Delta_{{\rm SUSYEW}}$ tends to a 
non-vanishing constant, and this constant tends to zero in the large
$M_{SUSY}$ limit. The solid lines in these figures 
correspond to the exact computation of the squarks/chargino loops, and
the dashed lines correspond to the results of the expansion as
in~(\ref{eq:amalitical}). Note that we have just considered the 
leading $\mathcal{O}(1)$ term in the expansion. 
The agreement between the exact results and the approximation derived
from the expansion is recovered when the second term in the
expansion is included.

The non-decoupling behavior emerging when the
SUSY mass scale is much larger than $M_A$, can be derived from the 
low-energy effective theory that is obtained 
by integrating out the SUSY particles.  This low-energy
effective theory contains two Higgs doublets, whose couplings to fermions
are  unrestricted ({\it i.e.}, each Higgs doublet couples to
{\it both} up-type and down-type quarks), characteristic of the
so-called general type-III two Higgs doublet model instead of the
type-II model that is assumed in the MSSM with no radiative corrections
included~\cite{carena}.

\begin{figure}\vspace*{0.2cm}
\epsfig{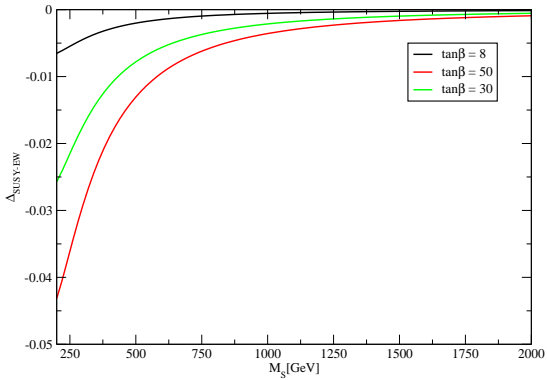}\vspace*{-0.6cm}
\caption{$\Delta_{{\rm SUSYEW}}$ as a function of $M_{SUSY}$ for
$M_{SUSY} = M_L=M_R=M_S = M_{\tilde g} = |\mu| = A_b = M_A$ and
$\tan\beta = 8$, 30, 50.}\vspace*{-0.6cm}
\label{fig:dec}
\end{figure}

The fact that decoupling is recovered when all SUSY mass parameters and
$M_A$ are equal is shown in Fig.\ \ref{fig:dec}. The  $\Delta_{\rm{SUSYEW}}$
corrections are plotted as a function of a common scale $M_S$ for
different values of $\tb$. Clearly, $\Delta_{{\rm SUSYEW}}$ decouples,
but this decoupling is delayed at large $\tan\beta$.  For example,
even at $M_{SUSY} = 1$~TeV,
$|\Delta_{{\rm SUSYEW}}| \simeq 0.5$\% for $\tan\beta \sim 30$.
The corrections can be $|\Delta_{{\rm SUSYEW}}| \simeq 2$\% for $\tan\beta
\sim 30$ and $M_{SUSY} = 250$~GeV.

\section{Conclusions}
\label{sec:conclusions}

In this paper we have studied the ${\mathcal{O}}(m_t^2)$ 
one loop Yukawa corrections to the
$h^0 b \bar b$ coupling, coming from diagrams involving the
exchange of virtual stops and charginos, 
in the limit of large SUSY masses.
We have performed expansions for the SUSY mass parameters large compared
to the electroweak scale. We demonstrate that in the limit of large
$M_A$ and large SUSY mass parameters, 
the corrections decouple like the inverse
square of these mass parameters, and
the SM expression for the $h^0 b\bar b$ one-loop coupling is recovered.
That is, the EW-Yukawa corrections to the $h^0 b \bar b$
coupling decouple in the limit of large SUSY masses and large $M_A$.
 However, if the mass parameters are not
all of the same size, then the decoupling behavior can be modified.  
If $M_A$ is light, then the corrections to the $h^0 b \bar b$ coupling
generically do not decouple in the limit of large SUSY mass parameters.

The decoupling behavior of the radiative corrections to the $h^0 b \bar b$
coupling implies that distinguishing $h^0$ 
from the SM Higgs boson will be very difficult if $A^0$
and the SUSY spectrum are heavy, even after one-loop SUSY corrections
are taken into account.  However, because
of the enhancement at large $\tan\beta$, 
the onset of decoupling is delayed, and the corrections can still
be at the percent level for $\tan\beta \sim 50$ and all SUSY mass
parameters and $M_A$ of order 1 TeV. 
Such effects could be detected in precision Higgs studies and provide
a critical window to the TeV-scale.


\end{document}